# Ultrahigh-Quality Magneto-Optical Resonances of Electromagnetically Induced Absorption in a Buffer-Gas-Filled Vapor Cell

Denis Brazhnikov[1,2], Stepan Ignatovich[1], Vladislav Vishnyakov[1],
Christina Andreeva[3], Vasiliy Entin[2,4], Igor Ryabtsev[2,4] and Mikhail Skvortsov[1]

[1]*Institute of Laser Physics SB RAS, 15B Lavrentyev ave., Novosibirsk 630090, Russia*
[2]*Novosibirsk State University, 2 Pirogova str., Novosibirsk 630090, Russia*
[3]*Institute of Electronics BAS, 72 Tzarigradsko chaussee blvd, Sofia 1784, Bulgaria*
[4]*Institute of Semiconductor Physics SB RAS, 13 Lavrentyev ave., Novosibirsk 630090, Russia*
Email: brazhnikov@laser.nsc.ru

**Abstract.** A new configuration for observation of magneto-optical subnatural-linewidth resonances of electromagnetically induced absorption (EIA) in alkali vapor has been verified experimentally. The configuration includes using two counter-propagating pump and probe light waves with mutually orthogonal linear polarizations, exciting an open optical transition of an alkali atom in the presence of a buffer gas. The main advantage of the novel observation scheme consists in the possibility of obtaining simultaneously high-contrast and quite narrow nonlinear signals. Here a 2.5-cm long rubidium-87 vapor cell filled with Ar buffer gas is used, and the excited optical transition is the $F_g=2 \rightarrow F_e=1$ of the $D_1$ line. The signals registered reach a contrast of 57.7% with a FWHM of 7.2 mG. The contrast with respect to a wide Doppler pedestal well exceeds 100%. To our knowledge, to date this is the best result for EIA resonances in terms of contrast-to-width ratio. In general, the results demonstrate that the new magneto-optical scheme has very good prospects for various applications in quantum metrology, nonlinear optics and photonics.

**Keywords:** *Coherent population trapping, electromagnetically induced absorption, buffer gas, vapor cell, level crossing*

**PACS:** 42.50.Gy; 32.80.Xx; 78.20.Ls

The quantum principles which form the basis of the operation of various quantum sensors yield remarkable advantages of these devices as compared with their classical counterparts. One of these key principles is based on the concept of coherent interaction between the electromagnetic fields and matter. This interaction can put matter into a special non-classical state, which can be used for accurate sensing of the physical fields, for precise control of light waves and other relevant applications. Therefore, this scientific direction attracts a great interest nowadays.

One of the most interesting technologies for the development of quantum sensors and optical controlling elements exploits the coherent population trapping phenomenon (CPT) [1,2]. For instance, many authors propose using CPT for sensing weak magnetic fields for medicine, biology and fundamental science (e.g., see [3–5]). CPT and related nonlinear effects can be also used to produce sensitive and compact optical switches and optical delay lines [6–9], which can be of principal importance for optical communications and quantum informatics. All-optical compact and chip-scale CPT atom clocks [10–12] can be also treated as quantum sensors of a microwave frequency.

In laser spectroscopy, the CPT regularly manifests itself as a steep decrease in the atomic vapor fluorescence or, equivalently, a steep increase in the transparency of the atomic vapor. This nonlinear effect is called electromagnetically induced transparency (EIT) [13]. The EIT is the operating principle of many technologies involved CPT. One of the most attractive features of the EIT resonances consists in their subnatural linewidth, which can be very narrow (from several hundreds of kHz and down to units of Hz) [14,15]. Also, a high EIT signal contrast of several to tens

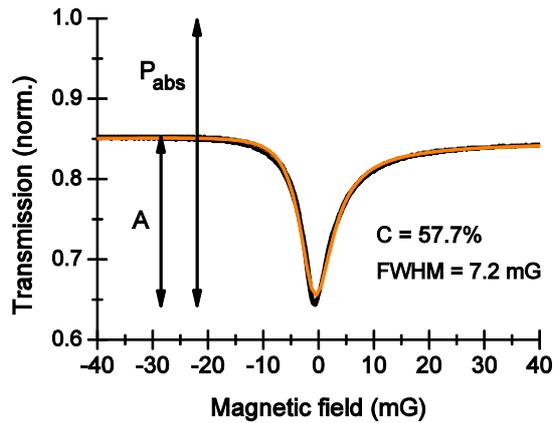

Figure 1. An example of the high-quality magneto-optical EIA resonance. Black curve is the experimental measurement of the transmitted probe-wave power normalized by the input power (3.5 μW). $P_{pump}$ = 390 μW. Orange curve is the Lorentzian fitting.

percent has been observed by many authors (e.g., see [16–23]).

The effect of electromagnetically induced absorption (EIA) was discovered about 20 years ago [24] and spectroscopically it looks as an opposite effect to the EIT. The EIA resonance manifests itself as a steep decrease in the cell transparency (see the example in Fig.1). Since its observation, the EIA effect has also found applications, like four-wave mixing [25], "fast" light experiments [26], magnetic field mapping of cold atomic samples [27]; still, until now it has not attracted as much attention as the EIT phenomenon. The main reason that seriously limits the area of EIA use, consists in the problem of obtaining a high-contrast and at the same time sufficiently narrow nonlinear resonance. Indeed, there are a lot of papers demonstrating a relatively good contrast of the EIA signals, but this is at the expense of large widths (for instance, see [28–33]). And, vice versa, the width can be as small as 10 kHz down to just several hundreds of Hz, but, unfortunately, these signals have an extremely low contrast (∼ 1 %) [34–36]. In work [18] EIA and EIT signals were observed for various atomic transitions in cesium and it can be seen that the EIA resonances have much lower contrast in comparison with the EIT signals. The first experiments on the EIA effect in a two-frequency and in the so called Hanle configurations also did not demonstrate any good contrast-to-width ratios [24,37].

Recently, interesting results were obtained in potassium atoms in a cell with antirelaxation coating of the walls [38]. The authors used an observation configuration consisting of two counter-propagating circularly polarized laser beams and a transverse magnetic field. Narrow EIA resonances at an unprecedented contrast up to 60% were obtained. However, the narrowest EIA signal observed in the paper had a width of about 66 kHz at a contrast of about 10%. Unfortunately, the higher contrast was measured for larger signal widths.

Summing up the aforesaid, we can state that the problem of obtaining a sufficiently large contrast-to-width ratio of EIA resonances has not been solved yet (with respect to real experiments). At the same time, it should be emphasized that the successful solution of this problem will promote the EIA resonances use in many directions of modern quantum technologies where EIT effect is actively involved at present.

In our previous work [39] we proposed to exploit two counter-propagating laser beams with mutually orthogonal linear polarizations to excite an open optical transition in an alkali-metal atom (Rb or Cs) for obtaining a high-quality (meaning high-contrast and narrow-linewidth) magneto-optical EIA signal. The proposed scheme of excitation allows one to use buffer gas to greatly improve the properties of the nonlinear resonances. Note that many other schemes do not allow using buffer gas to enhance the EIA resonance quality due to the rapid destruction of the excited state anisotropy. Still, no experimental verification of the new scheme or an in-depth theoretical description have been demonstrated. Here we partly eliminate this lack of knowledge and show the results of our experiments.

The experimental setup is schematically depicted in Fig, 2. The cell used contains isotopically enriched $^{87}$Rb vapor and around 12 Torr of Ar buffer gas. It has a cylindrical shape with length 25 mm and diameter 25 mm. The cell was irradiated by means of a DFB laser whose output light was split into two beams forming the counter-propagating pump and probe light fields. The laser frequency was locked to the center of the Doppler profile of the probe-wave absorption at the $F_g=2 \rightarrow F_e=1$ optical transition of the $D_1$ line. Magnetic field parallel to the laser beam propagation direction was produced by means of a solenoid and its value was scanned around zero value, yielding resonances of EIA. The experiments were carried out at 50°C temperature of the cell walls. The cell was heated by means of a special thermochamber made of copper. The laser beam diameter in the cell was around 5 mm. The probe-wave power was fixed to 3.5 μW, measured after the cell when the laser radiation was far from the one-photon resonance condition. A three-layer μ-metal magnetic shielding was used to decrease the influence of stray magnetic fields.

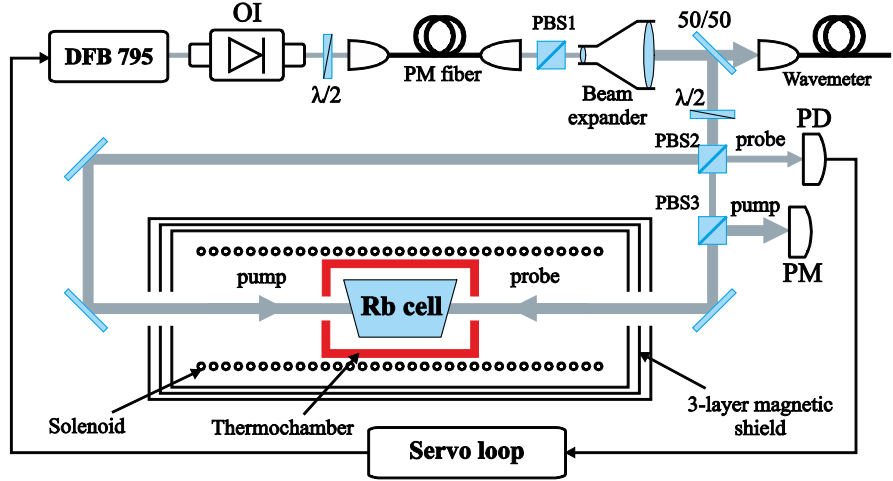

Figure 2. Sketch of the experimental setup: DFB 725 – distributed feedback diode laser with wavelength ≈ 795 nm and 1 MHz spectral bandwidth, OI – Faraday optical isolator, PM fiber – optical polarization-maintaining fiber, PBS – polarizing beam splitter, PD – photodetector, PM – power meter.

Fig. 3 shows the power broadening of the EIA resonance. It is seen that a FWHM as low as ≈ 2 mG was obtained. The value of the width at low pump-wave power levels can be affected and limited by power of the probe beam as well as the stray transverse magnetic field. The figure also contains the corresponding values of the Zeeman shift of the magnetic sublevels, which can be calculated using the formula: $\Omega \approx 0.73 \times B$ with $\Omega$ being the Larmor frequency (kHz) and B – the longitudinal magnetic field (mG). Therefore, a width as low as ≈ 1.5 kHz can be achieved.

There are several ways to define the contrast of a nonlinear resonance in the transmitted signal. For example, we can define it as the ratio between the EIA amplitude (A) and the total absorbed probe-wave power ($P_{abs}$) in the cell:

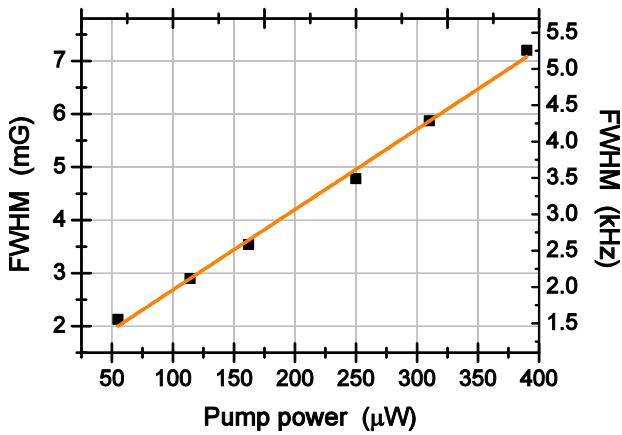

Fig. 3. Full width at half maximum of the EIA resonance vs. pump-wave power. Corresponding Zeeman splitting of magnetic sublevels is shown on the right vertical axis. Linear approximation is used.

$$C = \frac{A}{P_{in} - P_0} \times 100\% = \frac{T_{back} - T_0}{1 - T_0} \times 100\%. \quad (1)$$

See also the Fig.1. Here $P_{in}$ is the power of the probe beam when it is out of the one-photon resonance with the atoms, while $P_0$ is the output probe power at the center of the resonance. Transmission coefficient can be introduced as $T = P_{out}/P_{in}$. Then $T_{back}$ in (1) is responsible for the level of the wide background (≈ 0.85 in Fig. 1). For instance, the same definition of the contrast was used in [38]. Fig. 4(a) reveals the square-root-like dependence of the contrast on the pump-beam power. In particular, 57.7% contrast corresponds to 7.2 mG width. This combination of contrast and width is quite good for EIA-type of subnatural-linewidth resonances. Moreover, we expect that a contrast of 70% and higher can be achieved by further increasing the pump power (our setup was limited at 390 µW pump-wave power).

However, in the majority of works the contrast implies just a ratio between the EIA amplitude and amplitude of a wide pedestal, which suffers Doppler and buffer-gas collisional broadening (see [22,23,40], for example). In this case we have:

$$C_{rel} = \frac{T_{back} - T_0}{1 - T_{back}} \times 100\%. \quad (2)$$

It is worth noting that in our scheme $C_{rel}$ can be dramatically high (see Fig, 4(b)). In principle, it can exceed 100%, since the wide background amplitude decreases with pump power increase ($\propto 1/P$). This happens mainly because of the openness of the transition, as more atoms are pumped onto the non-resonant hyperfine level ($F_g = 1$ in our case) when the power is increased. This leads to a low absorption at the "wings" of the profile (see Fig. 1). The behavior of the resonance center is more complicated and it depends on several factors. In the simplest case when the stray transverse magnetic field is negligibly small, the atoms at the resonance center (B≈0) tend to stay on the $F_g=2$ level long time due to the CPT phenomenon, since they almost do not absorb the pump light (the probe beam power is too small to destroy or influence significantly the CPT state). Therefore, when the pump power increases, the background level of absorption decreases more rapidly than the level of probe-wave absorption at the center of the resonance, leading to ultrahigh value of the relative contrast. Of course, the existence of stray magnetic fields influences the CPT state created by the pump wave at B=0 and thus, the EIA contrast and width. Therefore, a high-quality magnetic shielding is quite necessary in order to obtain high-quality magneto-optical EIA resonances.

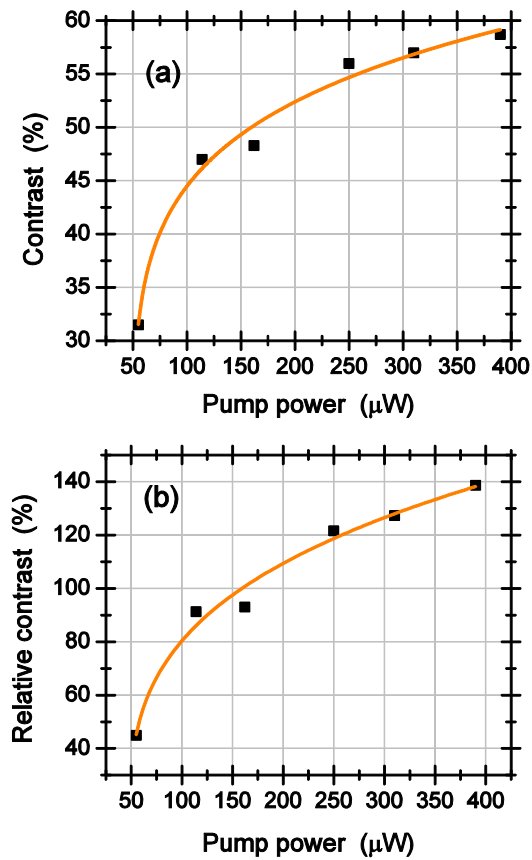

Fig. 4. (a) Contrast of the EIA resonance with respect to the total absorbed probe-wave power in the cell at B=0 and (b) contrast with respect to the amplitude of the wide background level. Nonlinear approximation is used.

The role of the atomic transition openness should be specially noted. Indeed, in many other schemes for observing EIA signals, the openness plays a very destructive role and prevents obtaining high contrasts (e.g., see [40–42]). However, for our scheme the openness of the transition is quite important for observing a high contrast.

The EIA contrast can reach really high values. To demonstrate this, we have performed numerical calculations based on the optical Bloch equations for the density matrix elements. Fig, 5 shows a contrast of about 90% can be obtained at relatively low pump beam power. It is also seen that the relative contrast can be equal to several hundreds of percent. The latter demonstrates a unique possibility for observing the probe-wave absorption profile composed of only subnatural-linewidth resonance, i.e. without any other broader structure as a pedestal. This is a feature of the EIA effect in our scheme of observation, even in comparison with the EIT resonances, which are subnatural-width structures superimposed on a broad background profile. Deviations of the theoretical values in Fig. 5 from the experimental ones (Fig. 4) can be explained by influence of the stray transverse magnetic field.

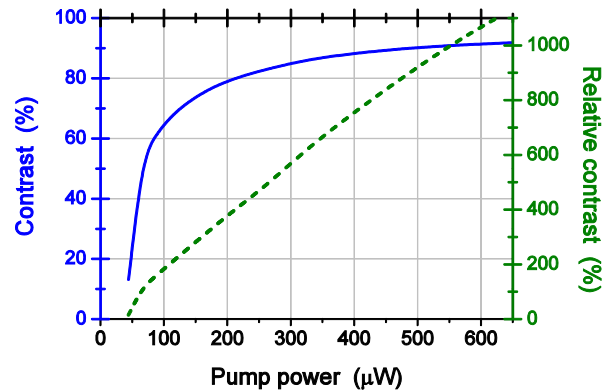

Fig. 5. Numerical calculations of EIA contrast vs. pump-wave power. Blue solid curve and left y-axis is for the contrast C (1), while green dashed curve and right y-axis stand for the relative contrast $C_{rel}$ (2).

To conclude, the work carried out has successfully proved the efficiency of the novel scheme for observation of high-quality magneto-optical EIA resonances. Moreover, we expect that further improvement of the experimental conditions (better magnetic field shielding, and use of smaller vapor cells to decrease the influence of the magnetic field inhomogenities, etc.) will lead to even better results.

ACKNOWLEDGMENTS

The work was supported by the Russian Science Foundation (17-72-20089) and Russian Presidential

Grant Program (NSh-6689.2016.2). The authors would like to thank Dr. Sanka Gateva and Dr. Latchezar Avramov from the Institute of Electronics (Sofia), Dr. Vitaly Palchikov from "VNIIFTRI" (Moscow region), Prof. Alexey Taichenachev and Dr. Alexey Lugovoy from ILP SB RAS for their assistance during our research.